\documentclass{article}

\usepackage{graphicx} 
\usepackage{dcolumn}  
\usepackage{bm}       
\usepackage{psfrag}   
\usepackage[mathscr]{euscript} 
\usepackage[numbers]{natbib}
\usepackage{url}

\begin{document}

\title{Constructing arbitrary Steane code single logical qubit fault-tolerant gates}

\author{Austin G. Fowler\\\\Centre for Quantum Computer Technology\\ School of Physics,
University of Melbourne\\Victoria 3010, AUSTRALIA.}

\date{\today}


\maketitle

\begin{abstract}
We present a simple method for constructing optimal fault-tolerant
approximations of arbitrary unitary gates using an arbitrary
discrete universal gate set.  The method presented is numerical and
scales exponentially with the number of gates used in the
approximation. However, for the specific case of arbitrary
single-qubit gates and the fault-tolerant gates permitted by the
concatenated 7-qubit Steane code, we find gate sequences
sufficiently long and accurate to permit the fault-tolerant
factoring of numbers thousands of bits long. A general scaling law
of how rapidly these fault-tolerant approximations converge to
arbitrary single-qubit gates is also determined.
\end{abstract}

In large-scale quantum computation, every qubit of data is encoded
across multiple physical qubits to form a logical qubit permitting
quantum error correction and fault-tolerant computation.
Unfortunately, only very small sets of fault-tolerant gates
$\mathscr{G}$ can be applied simply to logical qubits, where
$\mathscr{G}$ depends on the number of logical qubits considered,
the code used, and the level of complexity one is prepared to
tolerate when implementing fault-tolerant gates. Gates outside
$\mathscr{G}$ must be approximated with sequences of gates in
$\mathscr{G}$. The existence of efficient approximating sequences
has been established by the Solovay-Kitaev theorem and subsequent
work \citep{Solo95,Kita97,Niel00,Harr02}. In this paper, we describe
a simple numerical procedure taking a universal gate set
$\mathscr{G}$, gate $U$, and integer $l$ and outputting an optimal
approximation of $U$ using at most $l$ gates from $\mathscr{G}$.
This procedure is used to explore the properties of approximations
of the single-qubit phase rotation gates $R_{2^{d}} =$
diag$(1,e^{i\pi/2^d})$ built out of fault-tolerant gates that can be
applied to a single Steane code logical qubit. The average rate of
convergence of Steane code fault-tolerant approximations to
arbitrary single-qubit gates is also obtained.

Section~\ref{Solovay:section:opt_approx} describes the numerical
procedure used to find optimal gate sequences approximating a given
gate. A convenient finite universal set of fault-tolerant gates that
can be applied to a single Steane code logical qubit is given in
Section~\ref{Solovay:section:gates}.
Section~\ref{Solovay:section:phase} contains a discussion of phase
rotations $R_{2^{d}}$ and their fault-tolerant approximations,
followed by approximations of arbitrary gates in
Section~\ref{Solovay:section:arbitrary}.
Section~\ref{Solovay:section:conc} summarizes the results of this
paper and their implications, and points to further work.

\section{Finding optimal approximations}
\label{Solovay:section:opt_approx}

In this section, we outline a numerical procedure that takes a
finite gate set $\mathscr{G} \subset U(m)$ that generates $U(m)$,
a gate $U \in U(m)$, and an integer $l$ and outputs an optimal
sequence $U_{l}$ of at most $l$ gates from $\mathscr{G}$
minimizing the metric
\begin{equation}
\label{Solovay:eq:dist} {\rm dist}(U,U_{l}) = \sqrt{\frac{m-|{\rm
tr}(U^{\dag}U_{l})|}{m}}.
\end{equation}
The rationale of Eq.~(\ref{Solovay:eq:dist}) is that if $U$ and
$U_{l}$ are similar, $U^{\dag}U_{l}$ will be close to the identity
matrix (possibly up to some global phase) and the absolute value
of the trace will be close to $m$. By subtracting this absolute
value from $m$ and dividing by $m$ a number between 0 and 1 is
obtained. The overall square root is required to ensure that the
triangle inequality
\begin{equation}
\label{Solovay:eq:triangle} {\rm dist}(U,W) \leq {\rm
dist}(U,V)+{\rm dist}(V,W)
\end{equation}
is satisfied. This metric has been used in preference to the trace
distance used in the Solovay-Kitaev theorem \citep{Kita97,Niel00},
as the trace distance does not ignore global phase, and hence leads
to unnecessarily long global phase correct approximating sequences.

Finding optimal gate sequences is a difficult task, and the
run-time of the numerical procedure presented here scales
exponentially with $l$.  Nevertheless, as we shall see in
Section~\ref{Solovay:section:phase}, gate sequences of sufficient
length for practical purposes can be obtained.

For a set $\mathscr{G}$ of size $g=|\mathscr{G}|$ and a maximum
sequence length of $l$, the size of the set of all possible gate
sequences of length up to $l$ is approximately $g^{l}$. For even
moderate $g$ and $l$, this set cannot be searched exhaustively. To
describe the basics of the actual method used, a few more
definitions are required. Let $G$ denote a gate in $\mathscr{G}$.
Order $\mathscr{G}$, and denote the $i$th gate by $G_{i}$. Let $S$
denote a sequence of gates in $\mathscr{G}$. Order the possible gate
sequences in the obvious manner and let $S_{n}$ denote the $n$th
sequence in this ordering. Let $\{S\}_{l}$ denote all sequences with
length less than or equal to $l$. Let $\{Q\}_{l'}, l'<l$ denote the
set of unique sequences of length at most $l'$. Naively,
$\{Q\}_{l'}$ can be constructed by starting with the set containing
the identity matrix, sequentially testing whether $S_{n}\in
\{S\}_{l'}$ satisfies ${\rm dist}(S_{n},Q)>0$ for all $Q\in
\{Q\}_{l'}$, and adding $S_{n}$ to $\{Q\}_{l'}$ if it does. A search
for an optimal approximation of $U$ using gates in $\mathscr{G}$
begins with the construction of a very large set of unique sequences
$\{Q\}_{l'}$.

The utility of $\{Q\}_{l'}$ lies in its ability to predict which
sequences in $\{S\}_{l}, l>l'$ do not need to be compared with $U$
to determine whether they are good approximations, and what the
next sequence worth comparing is. To be more precise, assume every
sequence up to $S_{n-1}$ has been compared with $U$. Let
$\{S_{n-1}\}$ denote this set of compared sequences. Consider
subsequences of $S_{n}$ of length $l'$. If any subsequence is not
in $\{Q\}_{l'}$, there exists a sequence in $\{S_{n-1}\}$
equivalent to $S_{n}$. In other words, a sequence equivalent to
$S_{n}$ has already been compared with $U$, and $S_{n}$ can be
skipped. Furthermore, let
\begin{equation}
S_{n}=G_{i_{N}}\ldots G_{i_{k+l'+1}}G_{i_{k+l'}}\ldots
G_{i_{k+1}}G_{i_{k}}\ldots G_{i_{1}},
\end{equation}
where $G_{i_{k+l'}}\ldots G_{i_{k+1}}$ is the subsequence not in
$\{Q\}_{l'}$.  Let $Q(G_{i_{k+l'}}\ldots G_{i_{k+1}})$ denote the
next sequence in $\{Q\}_{l'}$ after $G_{i_{k+l'}}\ldots
G_{i_{k+1}}$. The next sequence with the potential to not be
equivalent to a sequence in $\{S_{n-1}\}$ is
\begin{equation}
G_{i_{N}}\ldots G_{i_{k+l'+1}}Q(G_{i_{k+l'}}\ldots
G_{i_{k+1}})G_{1}\ldots G_{1}.
\end{equation}
The process of checking subsequences is then repeated on this new
sequence. Skipping sequences in this manner is vastly better than
an exhaustive search, and enables optimal sequences of interesting
length to be obtained. It should be stressed, however, that the
runtime is still exponentially in $l$.

Highly non-optimal but polynomial runtime sequence finding
techniques do exist \citep{Kita97,Niel00,Kita02,Daws04} but will not
be discussed here.

\section{Simple Steane code single-qubit gates}
\label{Solovay:section:gates}

For the remainder of the paper we will restrict our attention to
fault-tolerant single-qubit gates that can be applied to the 7-qubit
Steane code. The Steane code representation of states $|0\rangle$
and $|1\rangle$ is \citep{Stea96}
\begin{eqnarray}
\label{Solovay:eq:zero_L} |0_{L}\rangle &=&
\frac{1}{\sqrt{8}}
         (|0000000\rangle + |1010101\rangle + |0110011\rangle \nonumber \\
&&\quad + |1100110\rangle + |0001111\rangle + |1011010\rangle \nonumber \\
&&\quad + |0111100\rangle + |1101001\rangle), \\
\label{Solovay:eq:one_L} |1_{L}\rangle &=&
\frac{1}{\sqrt{8}}
         (|1111111\rangle + |0101010\rangle + |1001100\rangle \nonumber \\
&&\quad + |0011001\rangle + |1110000\rangle + |0100101\rangle \nonumber \\
&&\quad + |1000011\rangle + |0010110\rangle). \\
\end{eqnarray}
An equivalent description of this code can be given in terms of
stabilizers \citep{Gott97} which are operators that map the logical
states $|0_{L}\rangle$ and $|1_{L}\rangle$ to themselves.
\begin{eqnarray}
\label{Solovay:eq:stabiliser1} \texttt{IIIXXXX} \\
\label{Solovay:eq:stabiliser2} \texttt{IXXIIXX} \\
\label{Solovay:eq:stabiliser3} \texttt{XIXIXIX} \\
\label{Solovay:eq:stabiliser4} \texttt{IIIZZZZ} \\
\label{Solovay:eq:stabiliser5} \texttt{IZZIIZZ} \\
\label{Solovay:eq:stabiliser6} \texttt{ZIZIZIZ}
\end{eqnarray}
States $|0_{L}\rangle$ and $|1_{L}\rangle$ are the only two that
are simultaneously stabilized by
Eqs~(\ref{Solovay:eq:stabiliser1}--\ref{Solovay:eq:stabiliser6}).

The minimal universal set of single-qubit fault-tolerant gates that
can be applied to a Steane code logical qubit consists of just the
Hadamard gate and the $T$-gate \citep{Niel00}
\begin{equation}
\label{Solovay:eq:Tgate} T = \left(\begin{array}{cc}
1 & 0 \\
0 & e^{i\pi/4} \\
\end{array} \right).
\end{equation}
For practical purposes, the gates $X$, $Z$, $S$, $S^{\dag}$ should
be added to this set, where
\begin{equation}
\label{Solovay:eq:Sgate} S = \left( \begin{array}{cc}
1 & 0 \\
0 & i \\
\end{array} \right),
\end{equation}
along with all gates generated by $H$, $X$, $Z$, $S$, $S^{\dag}$.
The complete list of gates that we shall consider is shown in
Eq.~(\ref{Solovay:eq:gate_set}). This is our set $\mathscr{G}$. Note
that gates $\{I,G_{1},\ldots,G_{23}\}$ form a group under
multiplication.

\begin{equation}
\begin{tabular}{rclcrcl}
$G_{1}$ &=& $H$ &\qquad\qquad\qquad& $G_{13}$ &=& $HS$ \\
$G_{2}$ &=& $X$ &\qquad\qquad\qquad& $G_{14}$ &=& $HS^{\dag}$ \\
$G_{3}$ &=& $Z$ &\qquad\qquad\qquad& $G_{15}$ &=& $ZXH$ \\
$G_{4}$ &=& $S$ &\qquad\qquad\qquad& $G_{16}$ &=& $SXH$ \\
$G_{5}$ &=& $S^{\dag}$ &\qquad\qquad\qquad& $G_{17}$ &=& $S^{\dag}XH$ \\
$G_{6}$ &=& $XH$ &\qquad\qquad\qquad& $G_{18}$ &=& $HSH$ \\
$G_{7}$ &=& $ZH$ &\qquad\qquad\qquad& $G_{19}$ &=& $HS^{\dag}H$ \\
$G_{8}$ &=& $SH$ &\qquad\qquad\qquad& $G_{20}$ &=& $HSX$ \\
$G_{9}$ &=& $S^{\dag}H$ &\qquad\qquad\qquad& $G_{21}$ &=& $HS^{\dag}X$ \\
$G_{10}$ &=& $ZX$ &\qquad\qquad\qquad& $G_{22}$ &=& $S^{\dag}HS$ \\
$G_{11}$ &=& $SX$ &\qquad\qquad\qquad& $G_{23}$ &=& $SHS^{\dag}$ \\
$G_{12}$ &=& $S^{\dag}X$ &\qquad\qquad\qquad& $G_{24}$ &=& $T$
\end{tabular}
\label{Solovay:eq:gate_set}
\end{equation}

We use this large set $\mathscr{G}$ as $H$, $X$, $Z$, $S$,
$S^{\dag}$ and their products can all be easily implemented with
transversal single-qubit gates. In contrast, the $T$-gate is
extremely complicated to implement \cite{Alif06}. Since we are
interested in minimal complexity as well as minimum length sequences
of gates in $\mathscr{G}$, it would be unreasonable to count
$G_{23}$ as three gates when in reality it can be implemented as
easily as any other gate $\{G_{1},\ldots,G_{22}\}$. Since
$\{I,G_{1},\ldots,G_{23}\}$ is a group under multiplication, minimum
length sequences of gates approximating some $U$ outside
$\mathscr{G}$ will alternate between an element of
$\{G_{1},\ldots,G_{23}\}$ and a $T$-gate. Note that the
$T^{\dag}$-gate is not required in $\mathscr{G}$ for universality or
efficiency as, in gate sequences of length $l\geq 2$, it is equally
efficient to use $S^{\dag}T$ or $TS^{\dag}$. The extra
$S^{\dag}$-gate is absorbed into neighboring $G_{i}$-gates, $i <
24$.

\section{Approximations of phase gates}
\label{Solovay:section:phase}

We now use the simple algorithm described in this paper to construct
optimal fault-tolerant approximations of phase rotation gates
\begin{equation}
R_{2^{d}}= \left(
\begin{array}{cc}
1 & 0 \\
0 & e^{i\pi/2^{d}}
\end{array}
\right).
\end{equation}
Gates $R_{2^{d}}$ are examples of gates used in the single-qubit
quantum Fourier transform that forms part of the Shor circuits
described in \citep{Fowl04b,Fowl03b}. Note that phase rotations of
angle $2\pi x/2^{d}$, where $x$ is a $d$-digit binary number, are
also required, but the properties of fault-tolerant approximations
of such gates can be inferred from $R_{2^{d}}$.

\begin{figure*}
\begin{center}
\begin{tabular}{cc}
\includegraphics[width=60mm]{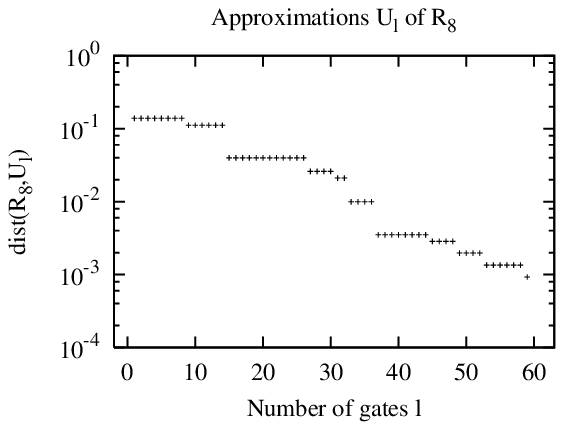} & \includegraphics[width=60mm]{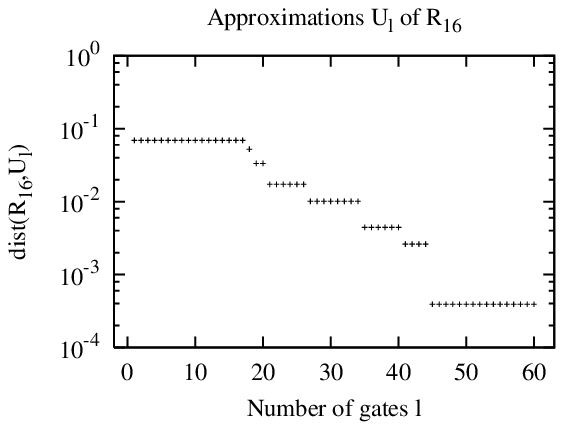}\\
\includegraphics[width=60mm]{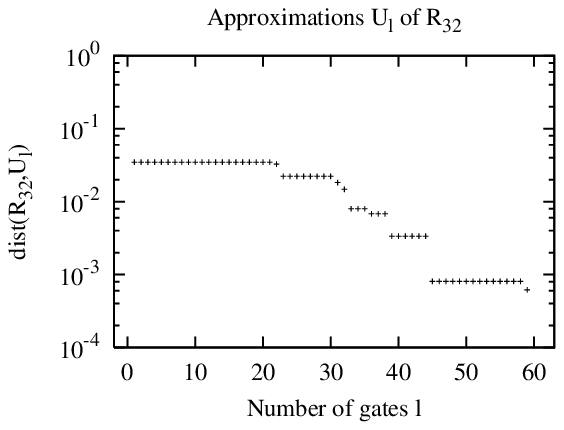} & \includegraphics[width=60mm]{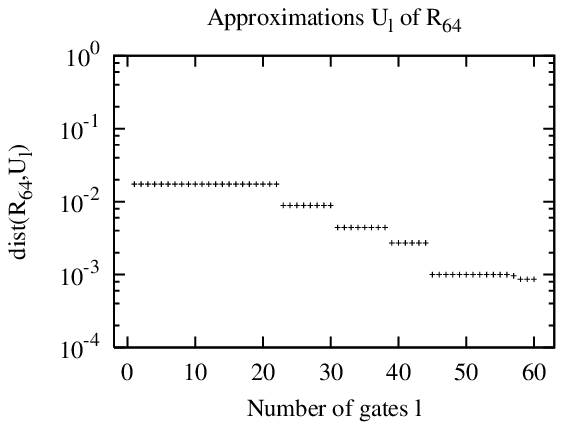}\\
\multicolumn{2}{c}{\includegraphics[width=60mm]{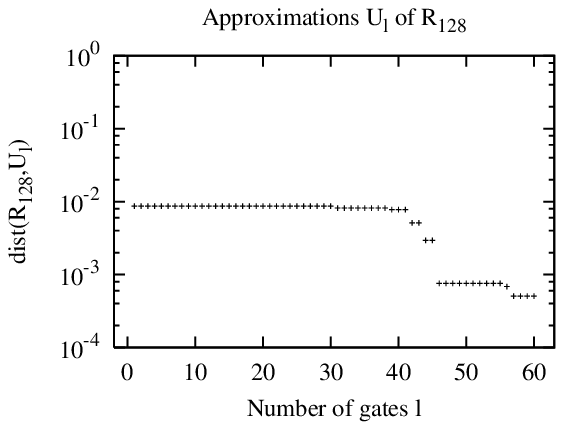}} \\
\end{tabular}
\end{center}
\caption{Optimal fault-tolerant approximations $U_{l}$ of phase
rotation gates $R_{2^{d}}$.} \label{Solovay:figure:constructions}
\end{figure*}

For a given $R_{2^{d}}$, and maximum number of gates $l$ in
$\mathscr{G}$, Fig.~\ref{Solovay:figure:constructions} shows ${\rm
dist}(R_{2^{d}},U_{l})$ where $U_{l}$ is an optimal sequence of at
most $l$ gates in $\mathscr{G}$ minimising ${\rm
dist}(R_{2^{d}},U_{l})$. For $d\geq 3$, $U_{1}$ is equivalent to
the identity. Note that as $d$ increases, $R_{2^{d}}$ becomes
closer and closer to the identity, lowering the value of ${\rm
dist}(R_{2^{d}},U_{1})$, and increasing the value of $l$ required
to obtain an approximation $U_{l}$ that is closer to $R_{2^{d}}$
than the identity. In fact, for $R_{128}$ the shortest sequence of
gates that provides a better approximation of $R_{128}$ than the
identity has length $l=31$. There are a very large number of
optimal sequences of this length. An example of one with a minimal
number of $T$-gates is

\begin{equation}
\label{Solovay:eq:U31}
\begin{tabular}{rcl}
$U_{31}$&=&$HTHT(SH)T(SH)T(SH)THTHT(SH)$ \\
       &&$THTHT(SH)THTHTHT(SH)T(S^{\dag}H)$
\end{tabular}
\end{equation}

Note that ${\rm dist}(R_{128},I)= 8.7\times 10^{-3}$ whereas ${\rm
dist}(R_{128},U_{33})= 8.1\times 10^{-3}$.  In other words
Eq.~(\ref{Solovay:eq:U31}) is only slightly better than the
identity. This immediately raises the question of how many gates
are required to construct a sufficiently good approximation.

In \citep{Fowl03b}, it was shown that
\begin{equation}
U=\left(
\begin{array}{cc}
  1 & 0 \\
  0 & e^{i(\pi/128 + \pi/512)}  \\
\end{array}
\right)
\end{equation}
was sufficiently close to $R_{128}$.  This is, of course, only a
property of Shor's algorithm, not a universal property of quantum
circuits. Given ${\rm dist}(R_{128},U)=2.2\times 10^{-3}$, a
fault-tolerant approximation $U_{l}$ of $R_{128}$ must therefore
satisfy ${\rm dist}(R_{128},U_{l})<2.2\times 10^{-3}$ to have a high
chance of being sufficiently accurate. The smallest value of $l$ for
which this is true is 46, and one of the many optimal gate sequences
satisfying ${\rm dist}(R_{128},U_{46})=7.5\times 10^{-4}$ is
\begin{equation}
\label{Solovay:eq:U46}
\begin{tabular}{rcl}
$U_{31}$&=&$HTHTHT(SH)THT(SH)T(SH)T(SH)THT$ \\
       &&$(SH)T(SH)THTHT(SH)T(SH)THT(SH)T$ \\
       &&$(SH)T(SH)THT(SH)THT(HS^{\dag})T$
\end{tabular}
\end{equation}
Note that implementing this long sequence of fault-tolerant gates
would necessitate the use of concatenation to ensure the inevitable
multiple errors during execution are reliably corrected.

\section{Approximations of arbitrary gates}
\label{Solovay:section:arbitrary}

\begin{figure}
\begin{center}
\includegraphics[width=7cm]{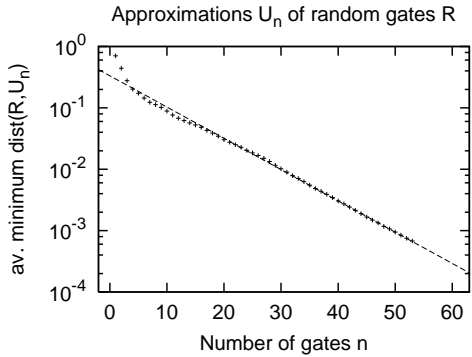}
\end{center}
\caption{Average accuracy of optimal fault-tolerant gate sequence
approximations of length $l$.} \label{Solovay:figure:average}
\end{figure}

In this section, we investigate the properties of fault-tolerant
approximations of arbitrary single-qubit gates
\begin{equation}
\label{Solovay:eq:arbitrary} U=\left(
\begin{array}{cc}
\cos(\theta/2)e^{i(\alpha+\beta)/2} & \sin(\theta/2)e^{i(\alpha-\beta)/2} \\
-\sin(\theta/2)e^{i(-\alpha+\beta)/2} &
\cos(\theta/2)e^{i(-\alpha-\beta)/2}
\end{array}
\right).
\end{equation}
Consider Fig.~\ref{Solovay:figure:average}.  This was constructed
using 1000 random matrices $U$ of the form
Eq.~\ref{Solovay:eq:arbitrary} with $\alpha,\beta,\theta$
uniformly distributed in $[0,2\pi)$.  Optimal fault-tolerant
approximations $U_{l}$ were constructed of each, with the average
${\rm dist}(U,U_{l})$ plotted for each $l$.  The indicated line
best fit has the form
\begin{equation}
\label{Solovay:eq:av_scaling} \delta=0.292\times 10^{-0.0511l}.
\end{equation}
This equation characterizes the average number $l$ of Steane code
single-qubit fault-tolerant gates required to obtain a
fault-tolerant approximation $U_{l}$ of an arbitrary single-qubit
gate $U$ to within $\delta={\rm dist}(U,U_{l})$.

\section{Conclusion}
\label{Solovay:section:conc}

We have described an algorithm enabling the optimal approximation of
arbitrary unitary matrices given a discrete universal gate set. We
have used this algorithm to investigate the properties of
fault-tolerant approximations of arbitrary single-qubit gates using
the gates that can be applied to a single Steane code logical qubit.
We have found that on average an $l$ gate approximation can be found
within $\delta=0.292\times 10^{-0.0511l}$ of the ideal gate. The
work here suggests that practical quantum algorithms should avoid,
where possible, logical gates that must be implemented using a
sequence of fault-tolerant gates since even the rotation gates used
in Shor's algorithm, which do not need to be implemented with great
accuracy, still require lengthy sequences. Quantum simulation
algorithms are expected to require far greater precision and thus
far longer sequences, and will be studied in future work.

\section{Acknowledgements}

We acknowledge support from the Australian Research Council, the
Australian Government, and the US National Security Agency (NSA) and
the Army Research Office (ARO) under contract W911NF-08-1-0527.

\bibliographystyle{unsrt}
\bibliography{../../References}

\end{document}